# Quantum communications infrastructure architecture: theoretical background, network structure and technologies.
# A review of recent studies from a European public infrastructure perspective

Adam M. Lewis[1] and Petra F. Scudo[1,2]

[1] *European Commission Joint Research Centre, Directorate E, Unit E.2 Technology Innovation in Security, TP 723 Via Enrico Fermi 2749, I-21027 Ispra (VA) Italy*
[2] *Fincons SPA, Via Torri Bianche 10, 20871 Vimercate (MB), Italy*
adam.lewis@ec.europa.eu; petra.scudo@ext.ec.europa.eu

**Abstract**
Progress in the development of techniques for the construction of multiuser quantum communications networks is reviewed in light of the plans for an EU quantum communications infrastructure (EU QCI). Quantum key distribution networks may be classified as trusted node; quantum repeater-entanglement-swapping based; point-to-multipoint based on passive beam splitters, active optical switchers or WDM; high dimensional-multipartite entanglement and flexible reconfigurable multipartite networks. Both satellite and terrestrial implementations are possible and ore both planned for EU QCI; their relative merits are discussed. Current technology falls short in several respects of what is required to address use cases well. Bit rates as a function of distance are currently limited by the characteristics of available devices and are an order of magnitude or more below the theoretical bounds. Non-ideal device behaviour can open loopholes for side-channel attacks. Limited range constrains network geometry. New techniques are being developed to overcome these problems. The more sophisticated schemes depend on ongoing advances in the generation and distribution of entanglement. Particularly promising are the measurement device independent protocol, which eliminates photon-detector related security weaknesses, and the twin-field protocol, which uses similar techniques to extend range, in principle even beyond previously accepted theoretical bounds. Agreement of keys between more than two users can be accomplished classically after pairwise quantum key exchange but direct quantum multipartite agreement using may offer advantages. The same applies to secret sharing. Practical quantum repeaters, to eliminate the need for trusted nodes, are closer to being realised due to recent progress in optical memories.

1. Introduction

The purpose of this paper is to review recent research advances in the field of quantum communication networks which are relevant to the feasibility and architecture of the proposed European Quantum Communication Infrastructure (Euro QCI) [1, 2, 3]. Our point of view is therefore that of government scientists working on plans for deployment of quantum communications technology at continental scale and we will focus on specific aspects rather than attempting a comprehensive general review of advances in quantum cryptography, such as [4].

The Euro QCI Declaration was signed by all European Union Member States, who "plan to work together for exploring how to make available across the EU an integrated Quantum-secure Communication Infrastructure, which would upraise Europe's scientific and technological capabilities in cybersecurity, quantum technologies and industrial competitiveness and its strategic autonomy". The Euro QCI aims to increase the security of sensitive data communication and storage by integrating quantum cryptography and quantum secure technologies in conventional communication networks, with priority to Government communication, critical infrastructures and data centre protection.

Euro QCI is intended to include both a terrestrial segment linking strategic national nodes and a space segment, known as SAGA (Security And cryptoGrAphic mission), which would be developed under the responsibility of the European Space Agency and consist of satellite quantum communication systems with pan-European reach. The first service would be the distribution of secret keys to encrypt classical communications (quantum key distribution, QKD) , and the long term goal a network to link future quantum computers at the quantum level, as the European section of a future global "quantum internet"[5].

High -level architecture studies and surveys of user requirements were conducted by European industry-academic consortia in 2020 and detailed design studies are now underway, funded by the European Commission and the European Space Agency. Plans for integration of national quantum networks in the Member States into the European infrastructure are included. All this work is complemented by testbed, research and development and standardisation projects, within the EU quantum technology programme.

For the purpose of establishing a European-scale secure network, choices of QKD hardware and protocol are dictated by cost, performance and security trade-offs and by the political goal of achieving strategic autonomy i.e. not to be overly reliant on imported technology. There is no a priori requirement to have a unique scheme throughout the network, providing that the devices are sufficiently interoperable. For example, one could imagine using a different protocol for short range links within cities and long range links between them.

In this paper, in relation to the planning of a large scale Euro QCI architecture, we review the most recent publications concerning quantum communication network structure for many simultaneous users. Aspects related to the use of more innovative protocols, entanglement based protocols, and on the length and accessibility of quantum links - considering both ground and space components are considered. Although EU QCI is initially being designed as a QKD network, research towards a quantum information network is being conducted by the Quantum Internet Alliance project of the EU Quantum Technology Flagship [6]. Attempts have been made to outline a roadmap for the evolution of quantum communication networks. S. Wehner, D. Elkouss and R. Hanson[7] identify six stages:

1. trusted repeater network
2. prepare and measure
3. entanglement generation
4. quantum memory
5. few qubit fault tolerant
6. quantum computing

The US Department of Energy "Quantum Internet Blueprint" [8], on the other hand, lists five milestones:

1. verification of secure quantum protocols over fibre networks,
2. inter-campus and intra-city entanglement distribution,
3. intercity quantum communication using entanglement swapping
4. interstate quantum entanglement distribution using quantum repeaters,
5. multi-institutional ecosystem between laboratories, academia, and industry to transition from demonstration to operational infrastructure

As will be shown here, significant progress is being made in these directions, driven by the need to overcome the limitations of existing quantum communications technology to provide a valuable QKD service, but with the long term vision in mind as well.

## 2. Channel capacity limits

End-to-end quantum channel capacities are limited by theoretical upper bounds, both in the absence and in the presence of quantum repeaters. Regarding the former, S. Pirandola, R. Laurenza, C. Ottaviani and L. Banchi [9] showed that performance of point-to-point protocols between two parties, without restrictions on local operations and classical communication, is limited by a general rate-loss upper bound providing the maximum theoretically achievable bit rate on a channel of given loss. This so-called PLOB bound represents the secret key capacity of point-to-point QKD, irrespective of the chosen protocol. It serves as a theoretical benchmark but rates achieved experimentally are, at best, an order of magnitude less, due to source and detector losses, fibre losses and other imperfections[10]. They cannot be surpassed by strategies involving local operations or classical communication but only by the introduction of quantum repeaters. In a subsequent paper, S. Pirandola derived upper bounds on the transmission of quantum information, entanglement and secret keys of arbitrary repeater-assisted channels[11] and quantum networks connected by arbitrary quantum channels. These upper bounds hold for both continuous variable (CV) and discrete variable (DV) systems, and under general loss-model assumptions. M. Ghalaii and S. Pirandola went on to devise a physical design for a CV memory-based system which would theoretically approach these capacities[12]. This is significant because CV QKD has the important advantage of not requiring single photon detectors, instead using homodyne detection of the quadratures with simpler and cheaper components, but it has not, in practical implementations, yet matched the performance of DV QKD.

PLOB-like bounds are extended to the domain of quantum networks under general communication strategies assumptions[13] (multiple-unicast, multicast, multiple-multicast) and for general quantum channels and protocols. These bounds limit the maximum rate at which parties can exchange quantum information as a function of the relative entropy of entanglement. An analogous bound has been recently derived for arbitrary network topologies for the specific case of single message multicast, of which the quantum conference key agreement protocol represents a special case (single-sender multicast). Recently, C. Harney and S. Pirandola[14] addressed the problem of optimal design of quantum network architectures that optimize end-to-end quantum communication applicable to the context of the quantum internet. They considered networks for which all nodes have the same number of neighbours, termed weakly regular graphs[15]. There is no reason, in principle, why a quantum network need satisfy this condition; it is a starting point for understanding how topology affects performance.

As remarked above, end to end capacities of quantum channels (and general routing quantum networks) connected by quantum repeater chains are, in principle, limited by a function of the relative entropy of entanglement. Today, though, the technology readiness is still very low, since quantum repeaters have only been demonstrated as laboratory devices, with very limited performance [4,16,17]. Moreover, with current technology the speed of long distance entanglement distribution links in fibre is limited. D. Aktas et al[18] achieved 1.1 bits/s over 150km , time entangled, in each of eight wavelength division multiplexed (WDM) channels. S. Wengerowsky et al [19] published results last year showing 1 bit/s, polarization entangled, over 192km . (For Euro QCI, the latter result is also significant because it was achieved in a submarine cable between Sicily and Malta, testing whether quantum communication over fibre could be an option even for island territories.)

Point-to-point channel capacity, with or without the assumption of multiband communication, can be compared with the capacity in the presence of repeaters. For distances above 20km, adding a certain number of repeaters allows a proportional increase in channel capacity.

The use of space based quantum links has been considered from the start as an invaluable resource to extend QKD channel distances and to connect separate ground-based networks. Simple schemes use the satellite as a trusted node, on the assumption that it is inherently less vulnerable than ground based nodes. But satellite links are also at the basis of higher-order, entanglement-based architectures for a global quantum information network[14, 20, 21, 22, 5]. The performance of a space-based quantum link depends critically on the optical link losses, which can be estimated using detailed models which take into account both free-space optical and atmospheric causes[24,25,26,27]. Applying the general PLOB bound for repeater-less quantum

channels, S. Pirandola and M. Ghalaii and S. Pirandola derive realistic rates for secret key and entanglement distribution through free space including losses due to diffraction, extinction, turbulence and pointing errors[28,29]. These techniques are then extended to study upper and lower bounds in both up-link and down-link satellite-ground QKD channels[30]. A secure key exchange requires a block of sufficient number of bits to be exchanged in a finite time, for space-based networks constrained by the satellite pass time. C. C. W. Lim et al have shown a security proof which is less demanding in block length than previous ones[31], making entanglement based QKD from space more feasible.

An improved hybrid system for integrated space-ground QKD has been recently implemented in China [32] updating the Chinese quantum backbone as well as the space-ground QKD link. Current secret key rate values for the Micius satellite space link are about 40 times higher than previously achieved ones, reaching 48kb/s. These results have been obtained by improving the source repetition rate (from 100MHz to 200MHz), the optical spatial mode matching and increasing the field of view of the optical receiving system. The system operates the BB84 decoy state protocol but is designed to be easily upgraded to both MDI[33,34] and TF[35,36] QKD, viewed as especially promising in ground based star networks to be integrated in the present backbone.

### 3. Multi-user quantum key distribution networks

QCI is expected to serve multiple users and, as such, to evolve into a complex topological structure, far surpassing the point-to-point networks which have been used for most QKD deployments to date. However, the possibilities for QCI to integrate multiple users is limited by intrinsic difficulties in achieving high performances over significant distances. Most QKD implementations involving multiple users are at metropolitan scale , with the important exception of the Chinese network[37]. Moreover, all have relied on trusted nodes, that is to say intermediate nodes where keys are stored as classical information, which therefore must be physically secured in some way to prevent eavesdropping. These are an inherent weakness which it is hoped to eventually eliminate, by means of quantum repeaters (when they become available).

That secret key rates (SKRs) are highly dependent on distance implies that a consideration of the topology of the network alone is insufficient to understand its expected performance. To date, networks have mostly been simple linear, rings or tree-like. In the latter, the secret key rate achievable on the main branches is divided up amongst the end points so it is critical that the main branches have adequate bandwidth, achievable either by using multiple dark fibres or by wavelength division multiplexing.

Multi-user quantum networks for QKD were introduced in 1997 by P. D. Townsend using optical fibres and BB84 [38]. The secret key rate for each user in Townsend's proof-of-principle passive optical network decays according to the number $N$ of users, proportional to $1/N$.

In recent years, several works have demonstrated the feasibility of multiuser QKD schemes, based on wavelength[39], space or time/code[40] or time/wavelength[41] division multiplexing or passive optical switching [42, 43, 44]. Switched networks are likely to be most appropriate for QKD in a metro setting, where flexibility to support many non-simultaneous users through software defined networking is more important than long-distance key rate.

Other schemes have introduced entanglement sharing[45],[46] to increase multipartite QKD efficiency, applying generalized BB84, BBM92 or E91 protocols.

If the stream of keys sent from one node is divided amongst many other nodes, then the SKR received by each will be divided by $1/(N-1)$. It is relevant to highlight in these respects that actual experiments involving multiuser QKD networks are at present characterized by very low secret key rates. A recent laboratory implementation based on multiuser MDI QKD with integrated photonic technologies on a star network reach records of a few tens of bits/s over 36dB attenuation [47]. (The authors say that this is equivalent to 180km, using the figure for fibre attenuation widely-used in simulations of 0.2dB per kilometer at 1550nm, but practical details of the splicing, connection and bending of cables usually make the field value worse). Over 100km, simulations demonstrate achievable key rates around 1kb/s based on actual parameters. W. Geng et al. demonstrate an achievable key rate of 87 kb/s over 20km fiber with integrated photonic technology and time-bin encoding [48]. These values are still very far from the theoretical limits predicted for both MDI QKD and TF QKD [49].

In general, QKD networks may be grouped in different configuration types – quantum repeater/entanglement swapping based networks; trusted node networks; point-to-multipoint networks – based on passive beam splitters, active optical switchers connecting two users at the time or WDM; high dimensional/multipartite entanglement networks; flexible reconfigurable multipartite networks allowing for simultaneous connection of every user to any other user.

A reconfigurable point-to-multipoint network, in which a user can be connected to any other user has been used to actively distribute entanglement within a multipartite static network [50]. In point-to-multipoint networks, as mentioned above for general multipartite networks, secret key rates as a function of users $N$ vary between $1/N$ [38] and an exponential decay in $N$ [41].

S. K. Joshi et al. [51] present an 8-user fully connected metropolitan network based on passive multiplexing, wavelength-dependent filters and beam splitters. In this network, a single source of polarization-entangled photons allows each user to simultaneously exchange a secure key with every other user. The experimental setup uses techniques from [50] and follows a scheme introduced in an earlier work [52], in which a multiple user network based on passive beam splitters distributes bi-partite polarization-entangled photons among 4 users and a single photon source signal is multiplexed into twelve wavelength channels. In S.K. Joshi et al. [51], the introduction of multiplexing and beam splitters causes an increase in loss for every user added to the network. As in the cases mentioned above, the secret key rate decreases with the number of users as 1/*N*. The authors highlight a dependence of the secret key rate on the network topology [53] and note that bi-partite entangled schemes outperform multi-partite ones whenever losses are significant, i.e. for distances of the order of 10km and above. Over such distances, secret key rates in [51] are of the order of a few kb/s. F. Appas et al. [54] use a complete graph to distribute bi-partite entanglement among *N* users, so that each user is connected to the remaining *N*-1. The entanglement source consist of an AlGaAs semiconductor chip which is used to implement the BBM92 protocol by means of DWDM and beam splitters. The work is considered to be an improvement of the scheme by C. Autebert et al. [55]. Secret key rate scales quadratically in *N*.

These and the above works suggest that bi-partite entanglement is probably the best performing protocol for multi-party QKD networks achieving bi-partite simultaneous quantum communication, when experimental resources are accounted for and over distances greater than 10km. M Epping et al. [56] investigate secret key rate as a function of fundamental network parameters and compare the use of bipartite with multipartite entanglement key agreement schemes. They show that multipartite entanglement sharing increases QKD secret key rates compared to bi-partite entangled schemes. The results are analytically derived in the limit of infinitely many protocol rounds and in the assumption of specific noise models.

Once keys have been agreed between all pairs of users, classical channels between them can be encrypted and used to distribute other keys. Therefore, if the only purpose of the network is key agreement, being limited to pairwise agreement is actually not very limiting. Multi-partite entanglement is necessary for schemes involving *N*-partite exchange of quantum information, and thus for quantum information networks requiring entanglement manipulation, storage and the use of quantum memories[56,57].

Multi user QKD networks, with a variability related to topology and protocol, allow for a simultaneous sharing by *N*=2 to 10 users over 20km links for secret key rates around 1kb/s. These values could be extended to up to 20 users once losses are reduced. Integrated photonic multiuser QKD networks running TF or MDI protocols reach distances of about 50km.

A recent experiment by M. Proietti et al. , based on polarization-encoded multipartite entangled GHZ (Greenberger-Horne-Zeilinger) states, achieves a four-party quantum key agreement protocol with a secret key rate of about 4b/s over 50km fibre [46].

Multipartite schemes exploiting entanglement sharing, in principle, could achieve higher secret key rates. However, the use of high dimensional multipartite entanglement is limited by the practical efficiency of preparing these states (see comment above on bipartite vs multipartite schemes). Entanglement sharing has also been used for quantum secret direct communication, i.e. using the channel directly to share the secret information, instead of distributing keys for encryption of classical communications. To achieve relatively reasonable data rates, maximally entangled EPR (Einstein-Podolsky-Rosen) pairs are used to densely encode two bits of classical information[58, 59]. In

a recent work, Z. Qi et al[60] demonstrate a 15-user quantum secure communication system based on 15 photon pairs via time division multiplexing (TDM) and dense wavelength division multiplexing (DWDM) on a 40-km fibre network.

Table 1. Different types of multipartite QKD networks and techniques

| Reference | system config. | topology | Multiplexing; Entanglement | Secret key rate (skr) | protocol | Status |
|---|---|---|---|---|---|---|
| G. Brassard et al., 2003 Ref.[39] | One-to-$N$ | tree | WDM | $K/N$ | BB84 | Theory plus feasibility experiments |
| M. Razavi, 2011 Ref.[40] | $N \times N$ | star | TDM/CDM (time/code division multiplexed) | decr Pol($N$) | BB84 | Theory |
| F. Grasselli et al., 2019 Ref.[61] | $N$ party entanglement sharing | star | Entanglement-based, W state | $K/N$-1 | Multiparty TF | Theory |
| P. Xue et al., 2017 Ref.[45] | $N$ party entanglement sharing | star | Entanglement-based, gen GHZ state | $K/N$ | Gen Ekert 91 | Theory |
| M. Proietti et al., 2020 Ref.[46] | $N$ party conference key | star | Entanglement-based GHZ state | decr Log($N$) | Gen Ekert 91 | Experiment |
| B. Fröhlich et al., 2013 Ref.[41] | One-to-$N$ | tree | Time and WDM | $K/N$-1 | Phase-encoding BB84 | Experiment |
| C. Ottaviani et al., 2019 Ref.[62] | $N$ party conference key | star with untrusted central node | CV MDI QKD Gaussian modulation | Fast decr. ($N$); low rates $N \geq 10$ | MDI | Theory |

| G. Murta et al., 2020 - Review[63] | N party conference key | star | | Comparing protocols skr(N) decr functions | 6 state-BB84; GHZ | Theory |
|---|---|---|---|---|---|---|
| C. Autebert et al. 2016 Ref. [55] F. Appas et al. 2021 Ref. [54] | NxN bi-partite entanglement sharing | Fully connected graph/mesh | DWDM + fibered polariz. beamsplitter | K/N^2 | BBM92 | Experiment Experiment |
| S. K. Joshi et al., 2020 Ref.[51] | N x N | tree | WDM and beamsplitter | K/N | BBM92 | Metro deployment |

### 4. Generation and distribution of multi-partite entanglement

The creation of multi-partite entangled photon states may be considered as a fundamental enabling technology for the realization of multiparty quantum communication networks. Traditionally, such states have been created by means of spontaneous parametric down-conversion (SPDC) in non-linear crystals. R. Prevedel et al. [64] reported generation by SPDC of Dicke states, a type of multipartite entangled state with completely undetermined spin direction in one plane[65], reaching fidelities with respect to ideal states of round 0.7 for 3-4 partite states and of about 0.5 for 6-party states. Multi-photon entangled states generated by SPDC with state of the art experimental equipment today reach a record of 12-photon entanglement, with fidelity 0.57 [66].

Fidelities of 0.78 have been reached with four-photon polarization-encoded quantum states via spontaneous four-wave mixing (SFWM) process in silicon spiral waveguides [67]. These settings are fully compatible with on-chip quantum processes [68], enabling lower costs and scalability. Optical integrated Kerr frequency combs [69] can be used to generate bi- and multi-photon entangled qubits[70] by SFWM with fidelities around 0.9 and 0.6 respectively. This method is also compatible with quantum memory set-ups and chip-based technologies.

SPDC and SFWM are spontaneous processes, so they make synchronization in entanglement detection more difficult. Consequently, schemes for the creation of multi-photon entanglement are being developed which are 'deterministic' rather than 'probabilistic', for example, using semiconductor quantum dots embedded in photonic crystal waveguides[71]. Fidelities of 0.9 may be achieved using chip-based quantum dot sources [72,73].

On-chip multi-partite entanglement and quantum teleportation have been achieved on a silicon chip by coherent control of an integrated network of nonlinear single-photon sources and multi-qubit entangling circuits. The production, control, and transceiving of states with entanglement fidelity approaching 0.9 took place on μm-scale silicon chips, fabricated by complementary metal-oxide-semiconductor processes suitable for mass production [74]. Time-bin entanglement using AlGaAs chips at room temperature has been achieved[55], also at fidelity 0.9 (inferred from stated visibility [75]). Hyper-entangled path-polarization 4 qubit cluster states on a photonic chip were realized with fidelities of about 0.85 [76].

5. Entanglement in space-based quantum communication and satellite-ground QKD

Given the limitations of entanglement distribution in fibre[18,19], satellite platforms are under consideration as an alternative for entanglement-based QKD, including for Euro QCI. But it is also challenging: to detect two entangled photons emitted by a satellite at two ground stations requires overcoming two sets of link-losses i.e. double the logarithmic (dB) attenuation.

Boone et al. proposed the first hybrid ground/space architecture [20]. Their scheme exploits entangled photon sources on low Earth orbit (LEO) satellites in combination with an array of quantum memories on ground nodes. This combination manages to increase entanglement distribution distances with respect to all schemes based exclusively on ground repeaters. Entanglement distribution rates are computed on a daily basis for different satellite orbits and compared with direct transmission from different orbits, mid-Earth orbit (MEO), specifically 2 000km, 10 000km, and geosynchronous Earth orbit (GEO).

S. Ecker et al. [25] have recently developed a quantum ground receiver for polarization-based QKD compatible with most existing optical ground stations with satellite tracking capabilities. The technology is tested to distribute entangled photons from an ultra-bright source over a 143-km free-space link with losses compatible with LEO down-links. The authors also propose an optimization model of satellite link parameters based on current channel and receiver conditions.

In any case, the use of hybrid geometries, with satellite + ground-based repeaters, allows for better entanglement distribution rates (except for low-medium orbits and ground distances below 4 000km). It should be noted however that optimal rates are achieved using LEO orbits and 8 repeater links. Optimal rates are about $10^6$ entanglement pairs per day, decreasing to less than 10 per day for distances of the order of 20 000km. Estimated entanglement distribution rates[30] derived from fundamental bounds should be considered as limits in the design of entanglement based space-ground links[77].

Furthermore, current state of the art finite key security analyses require blocks of the order of $10^4$

bits to achieve positive secret keys. This is particularly challenging in the case of entanglement based QKD over space links. A recent security analysis addressing optimization of block lengths to enable the feasibility of entanglement based space QKD has been published, for the case of the BBM92 protocol [31]. Transmission time window optimization and secret key length parameter dependence for the case of BB84 downlink are analyzed in [78].

6. Measurement device independent and twin-field QKD

These two protocols exploit entanglement in order to improve certain aspects of conventional QKD. MDI[33,34] QKD was conceived to overcome security assumptions on the detection phase while, at the same time, improving on the rates obtainable by using purely device independent solutions. In MDI QKD, the two trusted parties, Alice and Bob, send independent quantum signals (single photon states) to an untrusted middle relay. The secret key is established by means of a Bell state measurement performed in the middle relay and subsequent (classical) communication of the measurement outcome. Such a measurement provides post-selected entanglement that can be verified by the two trusted users and achieves a time-reversed entanglement protocol with many similarities with quantum teleportation, without the need for entangled photon sources. In an experimental scenario, the incoming signals are mixed in a 50:50 beam splitter, the outputs processed by two polarizing beam splitters and finally detected by two pairs of single-photon detectors. . The central-measurement-node geometry of MDI QKD is suitable for scaling to a multiple user switched network, in a star configuration[79].

Current distance records for MDI QKD are 404km in ultra-low loss fibre and 311km in standard fibre [80]. Non-ideal Bell state measurements and noisy channels do limit the performance in practice. At such distances, key rates are very limited, of the order of $10^{-2}$ bits/s. The same system allows, however, key rates of the order of a few kbits/s in the range of 100-150km, qualifying as a promising candidate for metropolitan scale networks. Performance values of in-field implementations of MDI QKD are still low and far from their theoretical limits, especially if one considers protocols which could theoretically exceed the PLOB bound. Recent in-field implementations based on MDI QKD with integrated photonic technologies reach a few tens of b/s over 180km link distances [47]. The chip in [47] works at 1.25GHz and realizes QKD with polarization encoding. R. I. Woodward et al. [81] introduce a GHz-clocked MDI QKD system based on directly modulated lasers to perform bit encoding. The system achieves key rates higher by about an order of magnitude compared to state of the art, specifically 1.97kb/s over 188km distances in fibre.

X. Wang et al. [82] present a feasibility assessment for space-based MDI-QKD based on the Micius satellite experimental parameters. The framework integrates orbital dynamics modelling and atmosphere losses to define optimal operational parameters, such as orbit height, elevation angle, transceiver apertures and atmospheric turbulence. Along with channel parameters, two-photon interference parameters for adequate implementation of the Bell state measurement in space are also considered. MDI QKD sources implement the decoy BB84 protocol using phase-randomized weak coherent pulses and transmit signals to an untrusted satellite based party who implements the Bell state measurement. Taking into account both experimental orbital parameters and realistic losses, the protocol achieves secret key rates as high as $1.9 \times 10^8$ bits per orbit.

MDI QKD over free space links was recently tested on a distance of about 20km in Shanghai [83], key rates of a few bits/s being achieved. The experiment may be considered as a step forward towards the implementation of an MDI QKD space link. In particular, the work enables an accurate estimate of the effects of atmospheric turbulence on the transmitted optical signals and thus allows the computation of time synchronization accuracy and frequency spreading.

The use of quantum memories on board a satellite in a configuration of uplink MDI QKD allows one to carry out the protocol exploiting the advantage of memory assistance. It should be noted that, in the absence of a space based quantum memory, the application of MDI QKD implies the realization of a Bell state measurement in the relay node using the two signal photons received by the ground links. An accurate estimation of the expected quantum bit error rate related to such a measurement under the assumption of realistic measurement devices in space environment and without quantum memories should therefore be performed.

M. K. Bhaskar et al. [84] report the experimental realization of a memory-enhanced MDI QKD network based on a single silicon-vacancy (SiV) colour centre integrated in a diamond nanophotonic cavity. The presence of the quantum memory in the intermediate node is used to overcome the limitations in the Bell state measurement generated by the lack of synchronization in the arrival of Alice's signal and Bob's signal. This enables a four-fold increase in the secret key rate of MDI-QKD over the loss-equivalent direct-transmission method while operating at megahertz clock rates. The memory-assisted rate also exceeds the PLOB bound for repeaterless communication.

W-F Cao et al.[85] introduce a flexible and extendable protocol named open-destination MDI, by combining open-destination teleportation [86] and MDI-QKD. This protocol allows the simultaneous establishment of a secure key by any two users in the network, thanks to the pre-sharing of an *N* partite entangled GHZ state. The flexibility of the scheme allows extension of the communication to a number *M* of the *N* users. The open-destination feature allows communication users not to be specified before the measurement step, which makes the network flexible and extendable. The network is to be compared with the establishment of $N(N-1)/2$ conventional MDI QKD links between any pair of parties.

Y. Fu et al. [87] propose combining decoy and MDI protocols to implement quantum conference key agreement and quantum secret sharing over distances of 190km and 130km respectively, as simulated with realistic experimental parameters. The method is based on the post-selection of entangled GHZ states and weak coherent pulsed sources and can be generalized to more than 3 parties.

A variant of MDI-QKD named Detector-Device-Independent (DDI) QKD[88] was conceived as a way to bridge MDI-QKD with conventional BB84 QKD, carrying the advantages of security against side channel attacks and exhibiting key rates close to the ones achieved by BB84 protocols[89]. The method is based on an MDI entanglement witness, which provides a condition on the security of the key against both external eavesdroppers and internal, dishonest, users. Although it does not quite succeed in eliminating the need to trust the Bell state measurement apparatus [90], with some extra monitoring measures it may be a good practical compromise. DDI QKD is introduced in [91] to achieve quantum secret sharing, secure against all detector side channel attacks, at up to 130km according to simulation.

In MDI multiparty quantum communication, the entanglement based generalised GHZ state measurement is performed by one of the communication parties or by a third party. The security of the scheme is based on measurement-induced entanglement, and thus the third party can be considered as an eavesdropper. T. Pramanik et al. [92] propose an equitable multiparty communication protocol where information balance among the communication parties is achieved without a third party. The protocol is based on the equal sharing of the GHZ state measurement by all communication parties.

There has been recent progress in implementing CV MDI QKD, however current distances reach only about 50km [93]. The use of post-selection to bridge the gap between DV and CV MDI protocols was introduced in this work, under the assumption of collective attacks, with the parties selecting only protocol instances that carry an advantage over the eavesdropper.

The Twin Field (TF) protocol (and its variants) are based on single photon interference, in a middle untrusted measurement station, of Alice's and Bob's phase-randomized optical fields[49]. The interference is detected by a single photon detector in the middle station and the outcomes are publicly announced. For reasonable distances, above 200km, TF QKD overcomes the PLOB bound for point to point repeater-less QKD because it simulates the presence of an active repeater (not an ideal one), allowing key rates between those of ideal point-to-point QKD and those of repeater-assisted QKD. The major technical barrier to the use of TF QKD is related to the phase drift experienced by the twin fields after travelling in long fibre links – which causes the protocol efficiency to drop drastically after about 500km of propagation.

F. Grasselli et al. [61] introduce TF QKD in a multipartite quantum conference key agreement scheme based on W state entanglement, with phase preselection[94]. Each party prepares an optical pulse in an entangled state with another qubit and transmits the pulse to the middle station, which applies a multi-port beam splitter and subsequently detects each event, announcing the corresponding results. In an honest implementation of the protocol, the parties distil the secret from their remaining qubits which constitute a W class state of $N$ qubits, a less entangled state than a generalized GHZ. The multipartite scheme key rates overcome direct bipartite transmission for increasing channel losses. We observe anyhow that, as the number of parties increases, the distilled secret key rate rapidly decreases. In a recent work, F. Grasselli et al. [95] prove that genuine multipartite entanglement is a necessary condition for perfect violation of Bell inequalities in DI multipartite QKD and conference key agreement schemes. S. Zhao et al.[96] introduce a phase matching quantum cryptographic conferencing based on post selection of GHZ states and twin field QKD. The protocol is immune to all detector side channel attacks.

In a recent in-field implementation, J.-P. Chen et al. [97] demonstrated TF QKD over 511km distance in ultra-low loss fibre, achieving a secret key rate which is three orders of magnitude higher than previous field tests indicate for this distance, and surpassing by ten times the absolute PLOB bound. This experiment is a step ahead towards the integration of point to point TF QKD inter-metro links within the existing Chinese backbone. More recently still, M. Pittaluga et al.[98] demonstrated TF QKD in 605 km of spooled fibre, using a dual-wavelength phase stabilization technique.

7. **Quantum repeaters**

Quantum repeaters, first introduced by H. J. Briegel, W. Dür, J. I. Cirac and P. Zoller[99] in 1998, in principle allow the extension of link distances by means of quantum memories, entanglement swapping and quantum error correction, to address the problem of link loss without resorting to classical trusted nodes, and lie at the basis of quantum internet networks[100]. Of these components, the most challenging are the quantum memories, for which several physical/chemical platforms have been investigated, including rare-earth-ion doped crystals, nitrogen-vacancy colour centres in diamond, Raman-addressed phonons in crystals, alkali metal vapours, molecules[101] and quantum dots[102].

The use of repeater chains naturalises transmission links to arbitrary quantum networks. Optimal performances of lossy quantum channels with different numbers of equidistant repeaters increase rapidly in performance for low numbers of repeaters, while saturating at around 100 consecutive repeaters[11]. M. Ghalaii and S. Pirandola proposed a continuous-variable, repeater-based communication network achieving the predicted theoretical upper bound[12]. J. Neuwirth et al. analysed prospects for memory-based repeaters exploiting semiconductor (GaAs) quantum dots[102] for the generation of polarization-entangled photons and their integration in photonic circuits. D. Miller et al. [103] explore the parameter regimes in which chains of quantum repeaters perform better in terms of secret key rates compared to direct point-to-point quantum communication. Their analysis includes quantum repeaters based on error-corrected high dimensional qudits (quantum computational units with more than two states).

The difficulty of constructing quantum repeaters, however, is comparable to that faced in building universal quantum computers. The complexity of the steps required means that their practical use, in terrestrial or space implementations, in a large operational network, like EU QCI, is only a long-term prospect. Quantum repeaters require the realization of multiple-step quantum communication processes[104]: the information stored in quantum states is transferred using photons by means of light-matter interfaces, which allows one to map stationary atomic or solid-state qubits into flying photonic qubits. Quantum state transfer may be achieved either via a deterministic or a heralded scheme (i.e. one where not every attempt to transfer quantum information is successful, but the successful cases are flagged). Heralded schemes comprise photon generation, measurement, and quantum teleportation (which requires both a classical and a quantum channel) and enable a faithful state transfer. Whereas quantum repeater nodes are essentially local quantum computers, quantum memories focus rather on information storage.

An intensive research effort is underway to make quantum repeaters practical and progress is encouraging. F. Kaneda et al. [105] demonstrate efficient synchronization of photons from multiple nonlinear parametric heralded single-photon sources (HSPSs), using quantum memories. The low-loss optical memories enhance the generation rate of coincidence photons from two independent HSPSs, while maintaining high indistinguishability of the synchronized photons. Each quantum memory triggered by a heralding signal from its corresponding HSPS, stores heralded photons for an arbitrary integer time until other sources produce pairs.

Rare-earth ions show particular promise as quantum memories because of their long coherence times and efficient storage within a solid-state platform[101]. A. Holzäpfel et al [106] presented an atomic frequency comb spin-wave memory using a europium doped yttrium orthosilicate crystal in a weak external magnetic field and dynamical decoupling (DD) which allows coherent optical storage times up to 530ms. This time has recently been dramatically increased to 1 hour by Y. Ma et al. [107] employing a zero-first-order Zeeman strong magnetic field. Coherence is verified via a time-bin interference experiment with a fidelity of 0.96. Neither result, however, was obtained at the single photon level, which would require improving the efficiency and signal to noise ratio.

Very recently, D. Lago-Rivera et al. and X. Liu et al. independently reported improved rare-earth memory repeater schemes in which the memory is a separate device from the heralded entangled photon source[16,17], whereas in previous schemes it has been embedded in it. The purpose of this modification is to give more flexibility in multiplexing the source without compromising its deterministic character. They demonstrated heralded entanglement distribution with this scheme, on laboratory-scale. The demonstration of Lago-Rivera et al. was at telecom wavelength.

NV diamond centres are also very promising platforms for memory. J-W. Ji et al. introduced a quantum repeater architecture based on non-cryogenic spin-photon interfaces and NV centres and quantum opto-mechanics, using a vibrating membrane, to bypass the need for a direct spin-photon interface, at room temperature[108]. The system's fidelity approaches 0.97 for time scales of the order of a few milliseconds. In recent work, an NV centre entanglement based network is used by M. Pompili et al. for the distribution of GHZ entangled states via a carbon 13 nuclear spin quantum memory, realizing a multi-node quantum network[100].

A completely different approach was taken in 2014 by K. Azuma et al.[109], who proposed an all-photonic scheme which does not require any matter-based quantum memory, or quantum error correction. They point out that for applications beyond QKD, which require complete sharing of a quantum state, some quantum memory, albeit shorter term, would still be required at the end points, it is only eliminated at the intermediate nodes. Y. Hasegawa et al. , with the authors of the earlier paper, [110] report a proof of principle experiment for a key component: an all-photonic time-reversed adaptive Bell measurement . The setup was tested on quantum teleportation. A complete all-photonic repeater has now been demonstrated by Z. Li et al.[111] using a 12 –photon interferometer to implement the repeater graph state (RGS). They conjecture that future systems might be hybrids in which RGS are used to reduce the requirements for quantum memory and quantum memory is used to simplify the RGS needed.

For the construction of a large quantum communications infrastructure, such as EU QCI, research and development on repeaters is of very great strategic importance. The implications of a successful quantum repeater would go far beyond the elimination of trusted-nodes, affecting network topology, the balance between space and terrestrial segments, network management, the required manufacturing capability and the standards, testing and certification processes to be followed. For the time being, the technology is not ready for deployment and which of the platforms being explored, if any, will emerge as the best remains unknown.

Meanwhile, research is already being conducted into the next steps, in preparation for when quantum repeaters become available. The first concept for a quantum network based on the distribution of graph states via quantum repeaters was proposed by M. Epping et al. [112] The introduction of graph states as distributed entangled resources allows the optimization of network structure and cost/performance factors in terms of graph properties. J. Wallnöfer et al. [113] introduce a quantum repeater network architecture for the distribution of multipartite (entangled) graph states based on the multipartite quantum hashing protocol. The scheme is analysed with respect to optimal scaling and allows for long distance communication and fidelities approaching 1. The multipartite approach is compared with the distribution of entangled pairs and shown to perform better storage-wise in certain parameter regimes and for general network topologies.

8. Quantum repeaters in space and satellite-ground QKD

Upper bounds on repeater-assisted ground based quantum networks also serve as a term of comparison for quantum communication over space links and for space based repeater-assisted channels[114, 115,]. In a recent work, C. Liorni, H. Kampermann and D. Brüss [114] compare the hybrid satellite-ground scheme [20] with a string-of-pearls repeater configuration, all space based. In their setting, quantum repeaters are placed in orbit and distribute photons by entanglement swapping to two distant users on the ground. The scheme exploits the lower losses of space links and thus allows for a decrease in the total number of repeaters. If entanglement distribution is used to perform QKD, one can show that the corresponding secret key rate is higher than what would be obtained both with a string of repeaters only on ground, or with the hybrid in-orbit/on-ground scheme described by K. Boone et al. [20]. Specifically, the use of space based repeaters increases the final key rate between two users, for distances below 6 000km, by as much as two orders of magnitude, achieving far better than the previous scheme for larger distances, where the hybrid orbit-ground scheme decays very fast. Depending on the number of repeaters used in orbit, SKRs could theoretically reach $10^5$ or $10^6$ bits/s for shorter links (below 2 000km), as compared to the hybrid scheme in[20], for which the maximum SKR would be between $10^3$ and $10^4$ bits/s.

These results, even though encouraging, should be interpreted in the light of state of the art quantum space-based technologies. Indeed, one should keep in mind that the implementation of quantum repeaters on board satellites introduces several technical challenges compared to ground based applications. However, the achievement of a Bose-Einstein condensate of rubidium atoms by NASA's Cold Atom Lab on board the International Space Station illustrates that a LEO microgravity environment can also have advantages for trapping topologies [116].

The advantage provided by quantum repeaters in space was recently exploited by M. Gündogan et al. [117] to analyse the performance of a memory assisted space-ground MDI QKD scheme. Memory assisted MDI QKD can be performed in both uplink and downlink configurations: in uplink, the Bell state measurement occurs between the quantum memories placed on a satellite acting as intermediate station; in downlink each quantum memory on the satellite sends a photon to the receiving ground station and the Bell state measurement occurs between the memories whose state is entangled with the one of the emitted photons. The simulations are compared with the experimental outcomes of the first Micius satellite experiment which is used as a baseline. With a source repetition rate of 20MHz (Micius used a source of 5.9MHz) and in the absence of quantum repeaters, the secret key rate approaches 1bit/s at 1000km ground distance. The model compares the performance of a standard E91 protocol without quantum repeaters, with the use of memory assisted entanglement, and predicts an increase up to two orders of magnitude in secret key rate between the two cases. It must be noted however that according to the simulation a simple increase in the source repetition rate leads to a significant increase in secret key rate even in the absence of repeaters. Indeed, the authors estimate that moving from 5.9MHz to 20MHz would lead to an increase of about an order of magnitude in secret key rate. In this specific context, even though the uplink configuration is characterized by higher losses, the downlink performance is limited by the classical communication necessary for the Bell state measurement and thus requires memories with longer storage times. The simulations also show that in order to achieve secret key rates higher by one or two orders of magnitude a large number of memories is required (up to 100) which would increase implementation and maintenance costs.

9. **Quantum conference key agreement**

It is interesting to point out that whereas the use of multipartite entanglement leads to higher secret key rates compared to non-entangled resources, it has been proved not to be a precondition for the achievement of an *N*-party conference key agreement.
Specifically, G. Carrara et al. [118] show that *N* parties may establish a secret conference key even when the state distributed in each round of the protocol is bi-separable (i.e. an ensemble of bi-partite entangled states). Instead, what is needed to establish a non-zero conference key is a set of measurements that detect entanglement across any partition of the set of *N* parties. This result is quite remarkable considering that entanglement is a precondition for two-party unconditionally secure QKD [119].

A comparison of the performance of bipartite vs multipartite entanglement-based conference key agreement is subtle and must consider the different resources required. In many applications, bipartite entangled states turn out to be more advantageous [51]. One can note that both protocols, bi-partite and multi-partite, require $N$-1 two-qubit gates for entanglement resource generation. A fully device independent (DI) $N$ party conference key agreement is analyzed by J. Ribeiro et al. [120,121] on the basis of an $N$-party Bell inequality. They show that its violation corresponds to a violation of the CHSH (Clauser-Horne-Shimony-Holt) inequality defined between one party and the other $N - 1$ and compare it with the use of $N$-1 individual QKD protocols. It is shown, that, for low noise regimes, the use of DI $N$-partite conference key agreement gives a better asymptotic key rate than separate DI QKD protocols. They include a security analysis of multiparty conference key agreement for general DI protocols based on a parity CHSH inequality check. Non-locality is certified via an effective Bell test of two parties depending on the measurement results of the remaining ones. T. Holz et al. [122] extend the results to generalized Bell inequality tests that are particularly suitable for DI QKD and DI conference key agreement protocols. These inequality tests allow one to increase secret key rates with respect to the CHSH parity check, especially in the presence of noise.

An interesting point deserving more attention is which classes of entangled states or, more in general, what degree of multipartite entanglement is necessary to implement full multi-party quantum networks. It is known that $N$-partite entangled states can be subdivided in different families corresponding to genuine $N$-partite maximally entangled states or combinations of ($N$-1) and lower degree of (non-maximally-entangled *N*-qubit) entangled states. General definitions related to entanglement measures for bipartite and multi-partite states can be found in [123,124]. An explicit derivation of these families for the case of 4-partite entanglement is provided by F. Verstraete, J. Dehaene and B. De Moor [125].

The estimation of required resources for *N*-party quantum protocols is of fundamental importance in optimizing cost/performance factors in quantum networks, given the challenges related to the generation and manipulation of multipartite entangled resources.

Whereas it is difficult to draw general conclusions from the results in *N*-party quantum networks, one can say nevertheless that, depending on network topology, protocol and multiplexing technique, the integration of multiple users in a QKD network decreases the secret key rate compared to point-to-point schemes and is higher for entanglement-based protocols.

How precisely protocol type, network topology, number of users and link distances affect secret key rates in multiuser QKD networks is currently a subject of investigation, and a systematic analysis of this problem with a general loss model is missing.

The derivation of generalized secret key rates and protocol dependencies in the context of complex multiuser quantum networks certainly deserves special attention in considering future multiuser architectures for QCI. To the best of our knowledge, multi-user quantum networks exhibit a general decrease in performances which ranges between exponentially decreasing secret key rates as a function of the number of users to rates decaying as $1/N$ or $1/(N-1)$ (using entanglement sharing, DI or TF schemes).

Multipartite QKD or conference key agreements have secret key rates decaying as $1/(N-1)$ or $1/N$ for optimized protocols and network geometries. Multipartite entanglement is not a precondition for $N$-party secret key sharing, therefore cost/benefit evaluations may prefer schemes that use linear combinations of 2(or more)-party entangled states. F. Grasselli, H. Kampermann and D. Brüss [61] show that Twin Field (TF) QKD can be generalized to a multipartite scenario achieving higher performance in terms of secret key rates compared to bipartite entanglement based schemes. Whereas other multipartite key agreement schemes achieve secret key rates decreasing as $1/N$, TF scales as $1/(N-1)$.

Graph states are multipartite entangled states which can be represented by mathematical graphs. They are naturally suited whenever entanglement sharing is required in general networks, themselves represented by mathematical graphs. As for the case of conference key agreements and secret sharing protocols, it is interesting to consider whether the sharing of a graph state is a fundamental or optimal resource for a given quantum protocol between users located in graph nodes. In the work mentioned above, J. Wallnöfer et al. [113] compare different network schemes based on quantum repeaters and the sharing of entanglement resources: bi-partite, multi-partite and hybrid entanglement sharing are compared in terms of fidelities and repeater storage. Storage requirements are higher for schemes based on bi-partite entanglement, however fidelity performances of the different schemes are comparable. Programmable 4-photon entangled graph states have been realized using a Si chip with fidelities of between 0.67 and 0.97, depending on the geometry of the state [126].

Cluster states are a special class of multipartite entangled states which arise in lattices with Ising type interaction (nearest neighbour plus uniform external field) ; they show improved robustness compared to GHZ or W states and can be generated using single-photon sources and entangling operations. Polarization-entangled linear cluster states can be generated via a single InGaAs quantum dot in an optical cavity with fidelities of about 0.5 for 4-photon entanglement [127].

Continuous variable, GHZ state schemes for multipartite QKD, conference key agreement and quantum secret sharing have also been devised, in both entanglement-based and prepare-and-measure versions, by Y. Wu et al. [128], using squeezed states, and by J. Zhou and Y.Guo [129], using coherent states. To overcome distance limits on these schemes, Z. Zhang et al. [130] propose a multipartite conference key network using CV GHZ states, with an untrusted source in the middle. This conference scheme is proved to be secure against coherent attacks. For less than $N$=10 users, the conference key rate is of the order of about 1kb/s on a range of 2km, reaching distances above 5km for 3 users. According to simulations, when the number of users exceeds 100, the scheme can still provide positive quantum conference keys to a radius of 500m. C. Ottaviani et al. [62] propose a modular CV network where each individual module is an MDI star network with users sending modulated coherent states to an untrusted central relay for conference key agreement. Multipartite secret correlations are created via generalized Bell detection. Different modules are connected through a classical one time pad channel so as to enable conference key agreement among the different users over large distances. Each module consists of a central (untrusted) relay performing a general *N*-mode Bell state measurement that allows an arbitrary number of users to share the same quantum conference key. Secret key rates decrease as a function of distance and decrease as the number of users increases, closely approximated by the scaling: maximum range = 2km/$N$, where *N* is the number of user in the star-network module. It is remarkable that this result, both for coherent and squeezed states, represents an improvement over DV MDI conference key networks, in general characterized by an exponential decay of the secret key rate in the number of users.

## 10. Quantum secret sharing

A related architecture option is the use of quantum secret sharing schemes. Secret sharing schemes in general are cryptographic threshold schemes ($k,n$) based on the subdivision of a secret data file into n pieces where k parties are capable of reconstructing the secret, while $k$-1 or fewer are not [131]. Their generalization into quantum secret sharing schemes was first introduced in 1998 by M. Hillery, V. Bužek and A. Berthiaume [132] for the case of three parties using entangled (GHZ) states and generalized to entanglement-based multiparty schemes by L. Xiao et al. [133] using generalized GHZ states. In parallel, quantum secret sharing schemes based on product states and realizable with state-of-the-art technology have also been proposed [57,134]. These protocols apply to classical secrets encoded in quantum states and achieve in principle up to 100% efficiency. They should be considered as an alternative to the application of classical secret sharing in combination with QKD and be compared to the latter both resource-wise and security-wise. Quantum secret sharing based on product states may be shown to be equivalent to $N$ independent BB84 QKD protocols and thus unconditionally secure [57,135], if the devices were ideal.

It is interesting to note that also in quantum secret sharing schemes the use of multipartite entanglement may not be necessary. Indeed G-P. Guo and G-C. Guo [57] argued that resorting to multipartite entanglement in the quantum secret sharing of classical information may prove counterproductive. Entanglement-based quantum secret sharing should thus be reserved for the sharing of quantum information, i.e. of quantum states. Two secret sharing schemes, one with and one without the aid of multipartite entangled states are discussed in terms of technology readiness level, implementation and scalability by C. Schmid et al. [136]. In particular, the use of 4-party multipartite entangled states is compared with the use of successive single photon states and local operations. The latter scheme is experimentally demonstrated for 6 parties and is easily scalable. A version of the scheme with improved security was proposed by G. P. He and Z. D. Wang [137].

## Conclusions

Quantum communication technology has been limited in range-speed performance and has been mostly confined to single links between two points, for the distribution of cryptographic keys which are then used for conventional classical communications. Large scale quantum networks were only possible with separately-secured classical "trusted-nodes", therefore not offering true end-to-end security. Current quantum networks also operate at secret key rates well below what is in principle possible. Intensive research is being conducted worldwide on protocols and multiplexing techniques to eliminate security weaknesses brought about by non-ideal device behaviour, to enable multiparty key agreement and to improve operational range and speed. Efforts to develop quantum repeaters to eliminate the need for trusted nodes are driving advances in quantum memories and the generation of entanglement. These are also the basic tools required for networks capable of communicating between quantum processors at the quantum level, so that techniques being realised now to improve QKD do indeed carry us forwards towards a "quantum internet".

## Acknowledgements

We would like to thank the following people who have commented on the manuscript and provided helpful suggestions:   A. Acin, M. Caleffi, M. Ghalaii, M. Gündogan, S. Joshi, S. Pirandola and Y.B. Sheng, and our colleagues I. Cerutti and M. Travagnin.

This work was conducted as part of the work programme of the European Commission's Joint Research Centre, project 30895.

**Data availability**

All the material used in this review is in the public domain, except Ref. 2 which is a limited distribution report, pending re-assessment, and Ref. 53 which is private correspondence, used with permission.

**Author contributions**

A M Lewis:  Introduction, conclusions, report structure, some of the analysis and a few references

P F Scudo:  Report structure, selection of references, most of the analysis

**Competing interests**

The authors declare no competing financial or non-financial interests.